\documentclass[twocolumn, 10pt, aps, superscriptaddress, floatfix, showpacs, prb, citeautoscript]{revtex4-1}
\usepackage{graphicx}
\usepackage{amsmath}
\usepackage{amssymb}
\usepackage{bm}
\usepackage{xcolor}
\usepackage[colorlinks, citecolor={blue!50!black}, urlcolor={blue!50!black}, linkcolor={red!50!black}]{hyperref}
\usepackage{bookmark}
\usepackage{tabularx}
\usepackage{mathtools}
\usepackage{microtype}

\DeclareMathOperator{\e}{e}

\DeclareMathOperator{\diag}{diag}

\newcommand{\vt}[1]{\mathbf{#1}}

\newcommand{\co}[2]{#2}
\renewcommand{\paragraph}{\co}

\DeclarePairedDelimiter\abs{\lvert}{\rvert}%
\DeclarePairedDelimiter\norm{\lVert}{\rVert}%
\makeatletter
\let\oldabs\abs
\def\abs{\@ifstar{\oldabs}{\oldabs*}}
\let\oldnorm\norm
\def\norm{\@ifstar{\oldnorm}{\oldnorm*}}
\makeatother

\newcommand{\ev}[1]{\langle#1\rangle}
\newcommand{\bra}[1]{\langle #1|}
\newcommand{\ket}[1]{|#1\rangle}
\newcommand{\bracket}[2]{\langle #1|#2\rangle}

\newcolumntype{L}[1]{>{\raggedright\arraybackslash}p{#1}}
\newcolumntype{C}[1]{>{\centering\arraybackslash}p{#1}}
\newcolumntype{R}[1]{>{\raggedleft\arraybackslash}p{#1}}

\graphicspath{{figures/}}

\begin{document}

\title{Single fermion manipulation via superconducting phase differences in multiterminal Josephson junctions}

\author{B. van Heck}
\affiliation{Instituut-Lorentz, Universiteit
  Leiden, P.O. Box 9506, 2300 RA Leiden, The Netherlands}
\author{S. Mi}
\affiliation{Instituut-Lorentz, Universiteit
  Leiden, P.O. Box 9506, 2300 RA Leiden, The Netherlands}
\author{A.R. Akhmerov}
\affiliation{Kavli Institute of Nanoscience, Delft University of Technology,
  P.O. Box 4056, 2600 GA Delft, The Netherlands}

\date{\today}

\begin{abstract}
  We show how the superconducting phase difference in a Josephson junction may be used to split the Kramers degeneracy of its energy levels and to remove all the properties associated with time reversal symmetry.
  The superconducting phase difference is known to be ineffective in two-terminal short Josephson junctions, where irrespective of the junction structure the induced Kramers degeneracy splitting is suppressed and the ground state fermion parity must stay even, so that a protected zero-energy Andreev level crossing may never appear.
  Our main result is that these limitations can be completely avoided by using multiterminal Josephson junctions.
  There the Kramers degeneracy breaking becomes comparable to the superconducting gap, and applying phase differences may cause the change of the ground state fermion parity from even to odd.
  We prove that the necessary condition for the appearance of a fermion parity switch is the presence of a ``discrete vortex'' in the junction: the situation when the phases of the superconducting leads wind by $2\pi$.
  Our approach offers new strategies for creation of Majorana bound states as well as spin manipulation.
  Our proposal can be implemented using any low density, high spin-orbit material such as InAs quantum wells, and can be detected using standard tools.
\end{abstract}

\maketitle
\section{Introduction}

\paragraph{Breaking TRS for single fermion levels is necessary for single spin manipulation and creation of Majorana fermions.}

In quantum mechanics, Kramers' theorem guarantees that in presence of time reversal symmetry the energy levels of a system with half-integer spin are doubly degenerate even if the spin rotation symmetry is broken \cite{Kramers1930,Wigner1932}.
A practical consequence of this theorem is that it is necessary to break time reversal symmetry in order to control single fermion states in a condensed matter system.
The energy separation of different spin states opens the way to spin detection and manipulation and is often a necessary element for spin qubits \cite{Loss1998} and spintronics \cite{Wolf2001,Zutic2004}.
The absence of Kramers degeneracy is also a fundamental requirement for the creation of unpaired Majorana bound states in topological superconductors \cite{Alicea2012,Beenakker2013review}.

\paragraph{Josephson rings focus effects of magnetic flux into a small area, and have other advantages, so they are better for breaking TRS than applying a huge magnetic field.}

\begin{figure}[t]
\centerline{\includegraphics[width=0.7\linewidth]{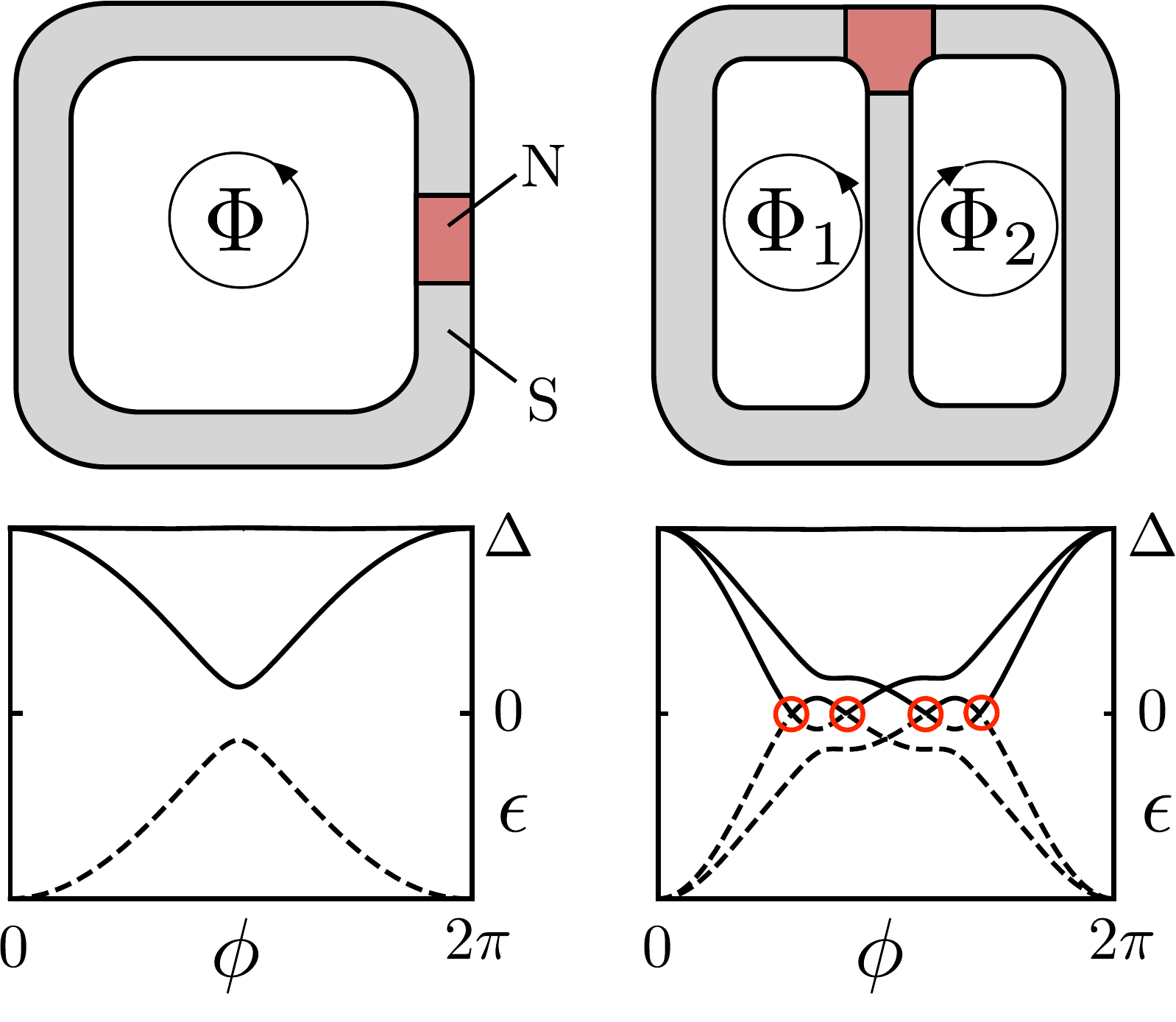}}
\caption{\emph{Top left}: A superconducting ring (grey) allows one to concentrate the effect of a magnetic flux $\Phi$ on the small area of a Josephson junction (red).
\emph{Bottom left}: The junction has subgap Andreev levels whose energy $\epsilon$ depends on the phase difference $2e\Phi/\hbar=\phi$.
Each level is doubly degenerate since in a short junction a finite phase difference does not induce a splitting of the Kramers degeneracy.
\emph{Top right:} As explained in this work, Kramers degeneracy can be efficiently removed in a three-terminal junction, even in the absence of an external magnetic field.
\emph{Bottom right:} Andreev spectrum for $2e\Phi_1/\hbar=-2e\Phi_2/\hbar=\phi$.
Both the splitting of Kramers degeneracy and Andreev level crossings at zero energy (marked by red circles) appear in the spectrum.}
\label{fig:josephson_loops}
\end{figure}

In order to provide fine-grained manipulation of electron states, a source of time reversal symmetry breaking should be local in space and easily tunable in time.
The superconducting phase difference across a Josephson junction satisfies these requirements.
It allows one to concentrate the effect of a magnetic flux penetrating a large superconducting ring into the small area of the Josepshon junction, whose spatial extent may be comparable to the superconducting coherence length $\xi$ (see Fig.~\ref{fig:josephson_loops}).
The magnitude of the energy splitting between a Kramers pair of bound states in the junction can then be comparable to the superconducting gap $\Delta$.
The magnetic field required to control the superconducting phase difference is rather small, and may be vanishing in the junction itself.
Flux bias loops applying this magnetic field allow one to address different Josephson junctions independently by tuning different fluxes, and have nanosecond response times.
These features seemingly make the superconducting phase difference the perfect source of time-reversal symmetry breaking for the manipulation of single fermion states.
In contrast, an external magnetic magnetic field seems to lose to phase differences in most respects: it needs to be a fraction of a Tesla to achieve a Zeeman splitting comparable to $\Delta$.
Such a field can only be tuned on the time scale of seconds and is rather hard to apply locally to only a part of a mesoscopic system.

\paragraph{However TRS breaking by a phase difference is suppressed in 2-terminal short
  JJs.}

Short Josephson junctions with Thouless energy $E_T$ much larger than the superconducting gap $\Delta$ are the most promising for single fermion manipulation, since they have the largest level spacing $\delta E \sim \Delta$.
Unfortunately, using phase difference as a source of time reversal symmetry breaking is ineffective in short two-terminal Josephson junctions.
This fact might seem surprising, since using symmetry considerations alone one would expect the spectrum of the Andreev bound states to be non-degenerate at a finite phase difference $\phi$.
As is well known, however, this expectation does not hold.
The Andreev energy levels $\epsilon_k$ are in one-to-one correspondence with the transmission eigenvalues $T_k$ of the scattering matrix of the junction in the normal state \cite{Beenakker1991}:
\begin{equation}\label{eq:2terminal_spectrum}
\epsilon_k=\pm\Delta\left[1-T_k\sin^2(\phi/2)\right]^{1/2}\,.
\end{equation}
In the absence of time reversal symmetry breaking in the normal state, the transmission eigenvalues $T_k$ are Kramers degenerate (see Ref.~\onlinecite{Bardarson2008} for a concise proof), and hence so are the Andreev levels.
Relaxing the short junction condition changes the scenario: spin-orbit coupling couples the spin of the bound states to the phase difference and lifts the Kramers degeneracy of the Andreev spectrum, albeit by a small amount of the order $\Delta^2/E_T$ \cite{Chtchelkatchev2003,Beri2008}.
Therefore, time-reversal symmetry can be broken only very weakly in a two-terminal junction.

\paragraph{We show that 3-terminal devices allow one to break TRS fully: there's large splitting of KD and GS fermion parity switches.}

In this work, we show how this serious limitation can be removed with a simple yet crucial change in the device geometry: the addition of an extra superconducting lead, as shown in Fig. \ref{fig:josephson_loops}.
Indeed, in devices with more than two superconducting terminals, the energy spectrum is not expected anymore to be in one-to-one correspondence with transmission eigenvalues.
We demonstrate that in this case the effect of time reversal symmetry breaking by superconducting phase differences alone leads to large splitting of the Kramers doublets comparable to the superconducting gap $\Delta$.
Naturally, since breaking the spin-rotation symmetry remains necessary, spin-orbit coupling is still an essential ingredient.
The non-degenerate Andreev spectrum makes these three-terminal junctions a promising platform for superconducting spin qubits \cite{Chtchelkatchev2003,Padurariu2010,Padurariu2012} and the creation of Majorana bound states, as we will discuss further in Sec.~\ref{sec:discussion}.

\paragraph{Among other things we find that a necessary condition for gap closing is a presence of a discrete vortex. It leads to a peak in the DOS at the Fermi level.}

As a consequence of the strong splitting of the Kramers degeneracy, crossings at the Fermi level can appear in the Andreev spectrum, corresponding to a switch in the ground state fermion parity \cite{Altland1997,Beenakker2013}.
We find that a necessary condition for the existence of a crossing at the Fermi level is the presence of a discrete vortex in the junction.
In other words, the gap in the Andreev spectrum can only close when the superconducting phases of the leads wind by $2\pi$ around the junction.
If this condition is satisfied, the spectral peaks in the density of states of the junction develop at the Fermi level as expected  \cite{Mehta,Altland1997,Ivanov2002,Bagrets2012} for a superconducting quantum dot with broken time-reversal and spin-rotation symmetries (symmetry class D of the Altland-Zirnbauer classification \cite{Altland1997}).

\section{General considerations}
\label{sec:analytics}

\subsection{Scattering formalism and bound state equation for multiterminal Josephson junctions}

\paragraph{We consider a Josephson junction with several leads coupled by a normal scattering region.}

\begin{figure}[htb]
\centerline{\includegraphics[width=0.7\linewidth]{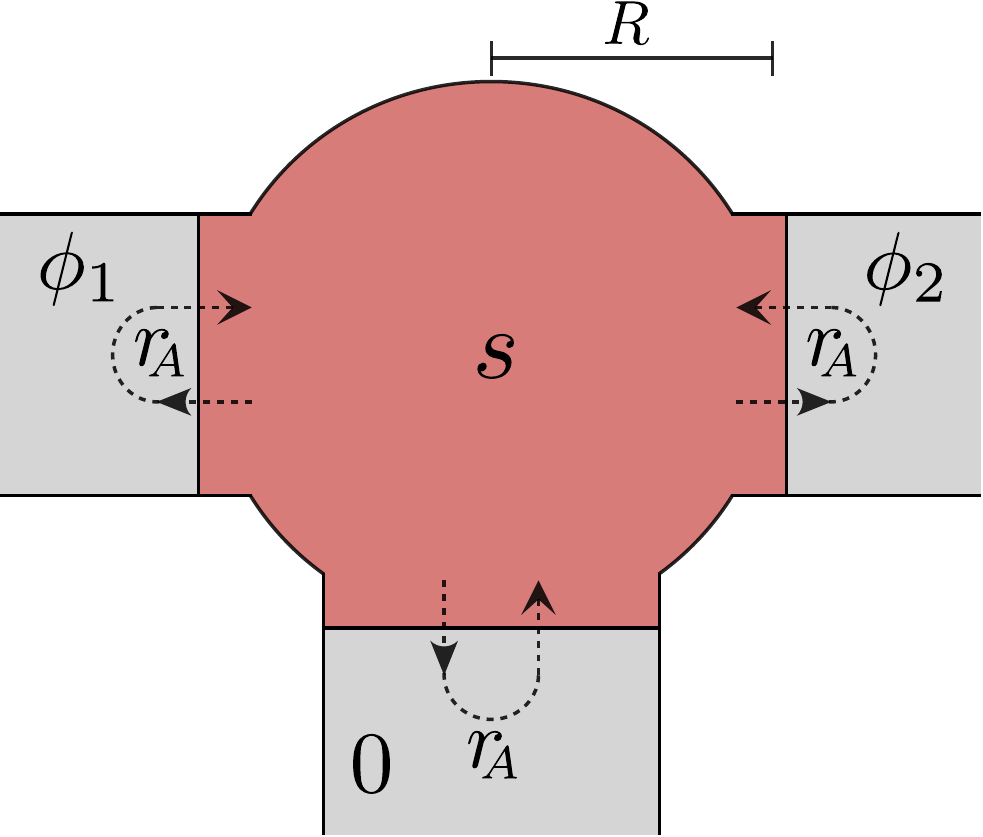}}
\caption{Three-terminal Josephson junction geometry. The scattering region (red) is a piece of a disordered two-dimensional material with spin-orbit coupling.
It is connected to three superconducting leads (grey).
In the normal state, the scattering region has a scattering matrix $s$.
At energies smaller than the superconducting gap $\Delta$, modes leaving the scattering region are reflected at the interface with the superconductor by Andreev reflection processes (black arrows), described by a scattering matrix $r_A$.}
\label{fig:dot}
\end{figure}

Three terminal Josephson junctions, such as the one shown in Fig.~\ref{fig:dot}, are the main focus of our work.
However, since most of our conclusions generalize naturally to the case of more terminals, we consider a junction with $m$ superconducting leads.
We assume that all of the leads have the same energy gap $\Delta$ and different phases $\phi_1, \dots, \phi_m$.
The coupling between the superconducting leads through the normal scattering region is fully characterized by the electron scattering matrix $s(\epsilon)$, with $\epsilon$ the excitation energy.
In general $s(\epsilon)$ is a $n\times n$ unitary matrix.
Its size $n=n_1+\dots+n_m$ is the sum of the number of incoming modes in the leads, counting spin.
The integers $n_1, \dots, n_m$ must be even due to the fermion doubling theorem \cite{Wu2006}.

\paragraph{Andreev bound states appear below the gap, and can be obtained from the scattering matrices.}

When $\abs{\epsilon} < \Delta$, an electron escaping the scattering region must be reflected back as a hole at the interface with the superconductor \footnote{In the presence of normal scattering in the superconductor, reflection from the superconductor can be represented as a combination of normal scattering followed by a subsequent perfect Andreev reflection. The former component can be combined with the scattering matrix of the normal region. Therefore, assuming perfect Andreev reflection does not reduce the generality of our results.}.
Closed trajectories of electron and hole superposition form Andreev bound states in the junction, which are confined by the superconducting pairing potential in the leads.
The spectrum of Andreev bound states can be expressed through two distinct scattering matrices: that of the scattering region $s_N$, and the scattering matrix $s_A$ describing Andreev reflection from a superconducting interface.
Both matrices are unitary and depend on the energy $\epsilon$.
As derived in Ref.~\onlinecite{Beenakker1991}, the condition for a presence of the bound state is given by:
\begin{equation}\label{bound_state_condition}
s_A(\epsilon)\,s_N(\epsilon)\,\Psi_\textrm{in}=\Psi_\textrm{in}\,.
\end{equation}
Here, $\Psi_\textrm{in}=(\Psi^e_\textrm{in}, \Psi^h_\textrm{in})$ is a vector of complex coefficients describing a wave incident on the junction in the basis of the modes incoming from the superconducting leads into the normal region.

\paragraph{Block structure of $s_N$ and $s_A$.}

Since in the normal region electrons and holes are not coupled, $s_N$ is block-diagonal in the electron-hole space.
We choose the hole modes as particle-hole partners of the electron modes and obtain
\begin{equation}\label{eq:sN}
s_N(\epsilon)=
\begin{pmatrix}
  s(\epsilon) & 0 \\
  0 & s^*(-\epsilon)
\end{pmatrix}.
\end{equation}
For more details regarding the relation between the basis choice for a scattering matrix and its discrete symmetries, see App.~A of Ref.~\onlinecite{Fulga2012}.
In the same basis, the Andreev scattering matrix $s_A$ is block off-diagonal since it couples only electron to holes and vice versa,
\begin{equation}\label{eq:sA}
s_A(\epsilon)=\alpha(\epsilon)\,
\begin{pmatrix}
0 & r^*_A \\
r_A & 0
\end{pmatrix}\,.
\end{equation}
The phase factor $\alpha(\epsilon)=\sqrt{1-\epsilon^2/\Delta^2}+i\epsilon/\Delta$ is due to the matching of the wave function at the interface between the normal region and the superconductors \cite{Beenakker1991}.

\paragraph{We can compute the energies as singular values of $r s - r s^T$.}

In the short junction limit, the energy dependence of the scattering matrix elements can be neglected,
\begin{equation}\label{eq:short_junction_condition}
s(\epsilon)\simeq s(-\epsilon)\simeq s(0)\equiv s\,.
\end{equation}
In that case the set of discrete Andreev levels $\{\epsilon_k\}$ can be computed by substituting Eqs.~(\ref{eq:sN},\ref{eq:sA}) into Eq.~\eqref{bound_state_condition} and solving the resulting eigenproblem for $\alpha\,$:
\begin{equation}\label{eq:alpha_eigenproblem}
\begin{pmatrix}
  s^\dagger & 0 \\
  0 & s^T
\end{pmatrix}
\begin{pmatrix}
  0 & r_A^* \\
  r_A & 0
\end{pmatrix}
\Psi_\textrm{in} = \alpha \Psi_\textrm{in}\,.
\end{equation}
It is convenient to apply to the above problem the Joukowsky transform
\begin{equation}
X\; \rightarrow\; -\frac{i}{2}\,\left(X-X^{-1}\right)\,,
\end{equation}
which maps $\alpha$ to $\epsilon/\Delta$. In this way, we obtain an eigenproblem directly for $\epsilon\,$:
\begin{equation}
  \label{eq:joukowsky}
  \begin{pmatrix}
    0 & -iA^\dagger\\
    iA & 0
  \end{pmatrix}
  \Psi_{\textrm{in}} = \frac{\epsilon}{\Delta}\, \Psi_{\textrm{in}},
\end{equation}
with
\begin{equation}
  \label{eq:adef}
A \equiv \tfrac{1}{2}\left(r_A s-s^Tr_A\right).
\end{equation}
Since $A$ is a normal matrix ($AA^\dagger=A^\dagger A$), its eigenvalues are equal to its singular values up to a phase, and as follows from \eqref{eq:joukowsky} its singular values are equal to $|\epsilon|$.
We now arrive at the simplified eigenproblem for the energies of Andreev levels:
\begin{equation}\label{eq:eig_equation}
A\,\Psi^e_\textrm{in} = \frac{\abs{\epsilon}}{\Delta}\,\e^{i\chi}\Psi^e_\textrm{in}\,
\end{equation}
The double degeneracy of the singular values of $A$ is a consequence of the fact that the eigenvalues of Eq.~\eqref{eq:alpha_eigenproblem} come in complex conjugate pairs, while only $\alpha$ with a positive real part are physical.
The reduction of the eigenproblem to the form of Eq.~\eqref{eq:eig_equation} is an important simplification which allows us to derive the properties of the Andreev spectrum of the junction.

\paragraph{TRS imposes constraints on $s$ and $r_A$.}

In the normal state the time-reversal symmetry is preserved in the junction and can be used to further constraint the scattering matrix $s$, which belongs to the circular symplectic ensemble\cite{Beenakker1997} (CSE, symmetry class AII).
Choosing a basis such that the outgoing modes are the time-reversed partners of the incoming ones results in $s$ becoming an antisymmetric matrix, $s=-s^T$.
Correspondingly, $A$ becomes the anticommutator of $s$ and $r_A$:
\begin{equation}\label{eq:anticommutator}
A=\tfrac{1}{2}\{s, r_A\}\,.
\end{equation}
Moreover, in the same basis in which $s$ is antisymmetric, the Andreev reflection matrix $r_A$ is diagonal,
\begin{equation}\label{eq:rA}
r_A=
\begin{pmatrix}
  i\e^{i\phi_1}\,\mathbf{1}_{n_1} & 0 & \dots & 0 \\
  0 & i\e^{i\phi_2}\,\mathbf{1}_{n_2} & \dots & 0 \\
  \vdots & \vdots & \ddots & \vdots \\
  0 & 0 & \dots & i\,\e^{i\phi_m} \mathbf{1}_{n_m}
\end{pmatrix}\,.
\end{equation}
We are now prepared to build a theory of multiterminal Josephson junctions.

\subsection{Kramers degeneracy splitting}

\paragraph{We first repeat the two-terminal case.}

For completeness, we first apply our formalism given by Eq.~\eqref{eq:eig_equation} to repeat the known result of the absence of the Kramers degeneracy splitting in two terminal short junctions.
For $m=2$, the Andreev reflection matrix $r_A$ has only two distinct eigenvalues $i\e^{i\phi_1}$ and $i\e^{i\phi_2}$, with multiplicity $n$.
\footnote{Generalization to unequal numbers of modes in two superconducting leads is straightforward, since in that case one of the leads will simply have several fully reflected modes leading to no extra Andreev bound states.}

\paragraph{The two-terminal result relies on polar decomposition of the scattering matrix.}

In this case, we can use the polar decomposition of $s$ \cite{Beenakker1997}:
\begin{equation}
\begin{pmatrix}
  U_1 & 0 \\
  0 & V_1
\end{pmatrix}
s
\begin{pmatrix}
  U_2 & 0 \\
  0 & V_2
\end{pmatrix}=
\begin{pmatrix}
  -\sqrt{1-T} & \sqrt{T}\\
  \sqrt{T} & \sqrt{1-T}
\end{pmatrix}\,.
\end{equation}
Here, $U_{1,2}$ and $V_{1,2}$ are $n\times n$ unitary matrices, while $T=\diag\,(T_1,\dots,T_{n_1})$ is a $n\times n$ matrix with doubly-degenerate transmission eigenvalues $T_k$ on its diagonal. Crucially, since
\begin{equation}\label{eq:rAcommutes}
\begin{pmatrix}
  U_{1,2} & 0 \\
  0 & V_{1,2}
\end{pmatrix}
r_A=r_A
\begin{pmatrix}
  U_{1,2} & 0 \\
  0 & V_{1,2}
\end{pmatrix}\,,
\end{equation}
the polar decomposition of $s$ carries on to $A$:
\begin{multline}
\begin{pmatrix}
  U_1 & 0 \\
  0 & V_1
\end{pmatrix}
A
\begin{pmatrix}
  U_2 & 0 \\
  0 & V_2
\end{pmatrix}=\\
\begin{pmatrix}
  -\sqrt{1-T}\e^{i\phi_1} & \tfrac{1}{2}\sqrt{T}(\e^{i\phi_1}+\e^{i\phi_2})\\
  \tfrac{1}{2}\sqrt{T}(\e^{i\phi_1}+\e^{i\phi_2}) & \sqrt{1-T}\e^{i\phi_2}
\end{pmatrix}\,.
\end{multline}
Diagonalization of the right hand side then immediately yields the spectrum of Eq.~\eqref{eq:2terminal_spectrum}.

\paragraph{The proof fails when $r_A$ has more than two distinct eigenvalues.}

It is easy to recognize that this derivation cannot be extended to the multiterminal case.
Indeed, if $r_A$ has more than two distinct eigenvalues, Eq.~\eqref{eq:rAcommutes}  does not hold anymore and there is no polar decomposition which can be simultaneously applied to both $s$ and $A$.
The correspondence between Andreev levels and transmission eigenvalues of $s$ is then lost.
As a consequence, we expect the spectrum of a multiterminal junction to consist of non-degenerate levels, unless the phases in the leads are tuned in such a way that the two-terminal case of only two distinct eigenvalues of $r_A$ is restored.

\paragraph{The Kramers splitting never exceeds the level spacing.}

If spin-rotation symmetry is strongly broken, and the phase differences are not small, there is no small parameter in the eigenproblem of Eq.~\eqref{eq:eig_equation} with more than two terminals.
This means that the energy splitting between Kramers partners becomes comparable to the Andreev level spacing in the junction, and scales as $\Delta/n$, the maximal possible value.
A simple estimate shows that, as one would expect, the splitting of Kramers degeneracy obtained using superconducting phase differences may never exceed the normal level spacing in the scattering region.
Indeed, for the junction to be in a short junction regime, $\Delta$ should be much smaller than the Thouless energy $n \delta_0$, with $\delta_0$ the normal level spacing in the scattering region.
This immediately gives an upper bound of $\delta_0$ on the Kramers degeneracy breaking.

\subsection{Lower bound on the energy gap and existence of zero-energy solutions}
\label{sec:lower_bound}

\paragraph{Also in a multiterminal junction we expect a lower bound for the energy.}

For the two-terminal case, Eq.~\eqref{eq:2terminal_spectrum} implies a lower bound $\abs{\epsilon}\geq \Delta \cos(\phi/2)$ on the energy of the Andreev states, irrespective of the junction details.
Inspecting Eq.~\eqref{eq:anticommutator}, we see that when all $\phi_i$ are close to each other, $r_A$ is an almost constant matrix, so that $\{s\,,r_A\}/2$ is almost unitary, and consequently all of the Andreev energies are close to $\Delta$.
This suggests that it is natural to expect some lower bound for $\epsilon$ also in the multiterminal case.

\paragraph{We reduce the problem of finding a bound wrt to a matrix to that wrt to two vectors.}

To determine this lower bound, we rewrite the eigenvalue equation \eqref{eq:eig_equation} as:
\begin{subequations}
  \begin{eqnarray}
    s\,r_A\ket{\Psi}+r_A\ket{\Psi'}=\frac{2\abs{\epsilon}}{\Delta}\,\e^{i\chi}\,\ket{\Psi}\,,\\
    \ket{\Psi'}\,\equiv s\,\ket{\Psi},\; \norm{\Psi} = \norm{\Psi'} = 1.
  \end{eqnarray}
\end{subequations}
The two above equations dictate that $s$ is a linear mapping such that
\begin{subequations}\label{eq:s_as_mapping}
\begin{align}
\ket{\Psi}&\,\xrightarrow{s}\,\ket{\Psi'},\\
r_A\ket{\Psi}&\,\xrightarrow{s}\,\frac{2\abs{\epsilon}}{\Delta}\,\e^{i\chi}\,\ket{\Psi}-r_A\ket{\Psi'}\,.
\end{align}
\end{subequations}
Since $s$ is unitary, these equations may be satisfied for given $\Psi$ and $\Psi'$ if and only if the scalar products between the vectors on the left and right hand sides of Eqs.~\eqref{eq:s_as_mapping} are preserved.
Hence, a necessary and sufficient condition for the existence of a solution is
\begin{equation}
\bra{\Psi}\,r_A\,\ket{\Psi}+\bra{\Psi'}\,r_A\,\ket{\Psi'}=\frac{2\abs{\epsilon}}{\Delta}\,\e^{i\chi}\,\bracket{\Psi'}{\Psi}\,.
\end{equation}
Taking the absolute value on both sides and using the Cauchy-Schwarz inequality $\abs{\bracket{\Psi'}{\Psi}}\leq \norm{\Psi'}\norm{\Psi}=1$ yields the lower bound
\begin{equation}\label{eq:lower_bound}
\abs{\epsilon}\geq\tfrac{1}{2}\,\Delta\,\abs{\bra{\Psi}\,r_A\,\ket{\Psi}+\bra{\Psi'}\,r_A\,\ket{\Psi'}}\,.
\end{equation}
We have thus reduced the problem of finding the lower bound with respect to a unitary matrix $s$ to a problem of finding the lower bound with respect to two vectors.

\begin{figure}[htb]
\centerline{\includegraphics[width=0.9\linewidth]{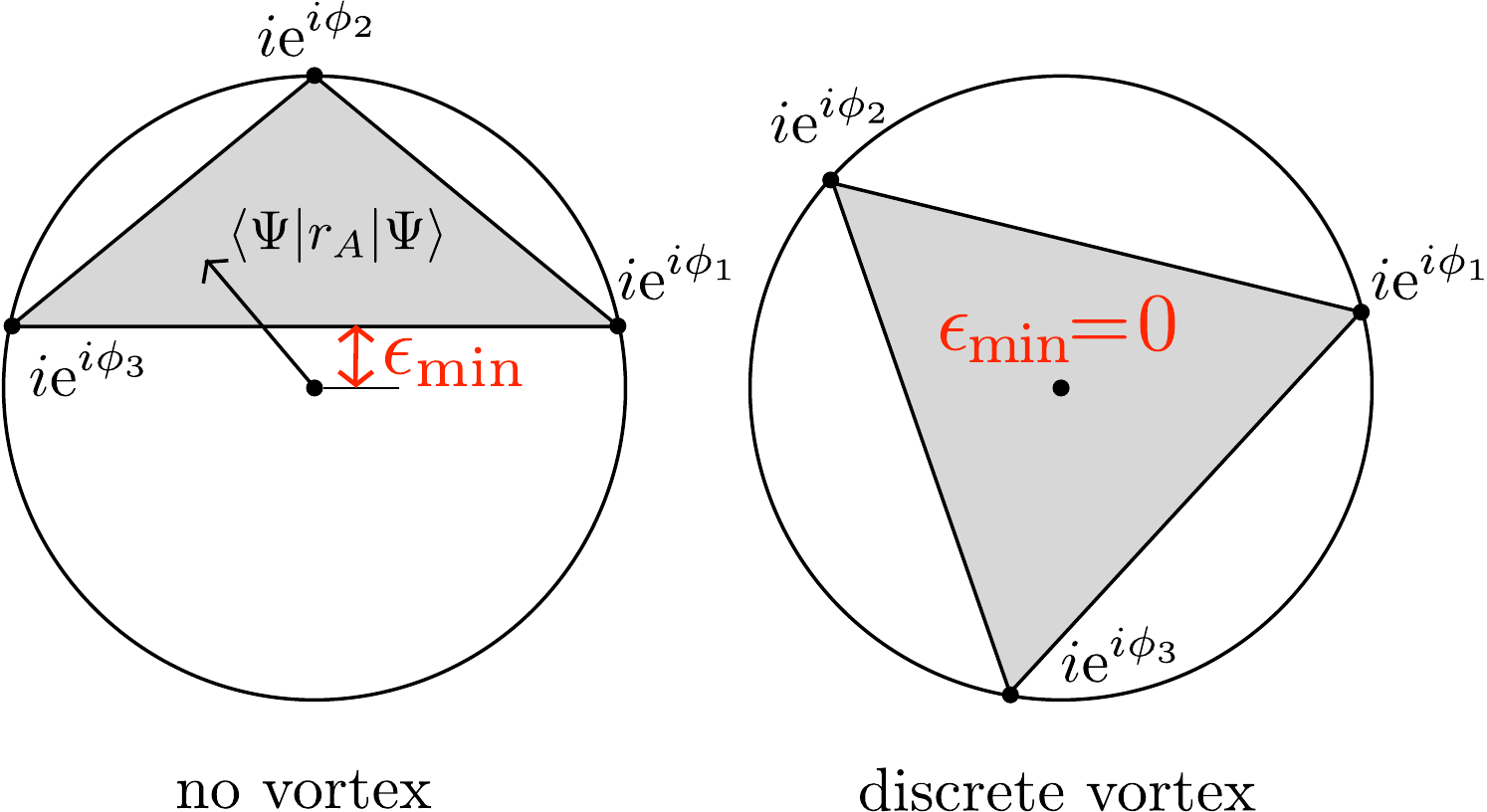}}
\caption{Geometrical illustration of Eq.~\eqref{eq:lower_bound} in the case of three leads. The sum of the scalar products $\tfrac{1}{2}\bra{\Psi}r_A\ket{\Psi}$ and $\tfrac{1}{2}\bra{\Psi'}r_A\ket{\Psi'}$ must lie within the triangle on the complex plane whose vertices are the eigenvalues $i\e^{i\phi_1}, i\e^{i\phi_2}, i\e^{i\phi_3}$ of $r_A$. In the left panel, these phases do not surround the origin and the lowest allowed energy (in units of $\Delta$) is the minimum distance between the polygon and the origin [Eq.~\eqref{eq:lower_bound2}]. In the right panel, the phases surround the origin, a discrete vortex is present in the junction and zero-energy solutions are allowed.}
\label{fig:geometrical_solution}
\end{figure}

\paragraph{The remaining problem has a nice geometrical solution.\\ }

The two scalar products in Eq.~\eqref{eq:lower_bound} are weighted sums of the eigenvalues of $r_A$ with total weight equal to one.
This means both these scalar products, as well as their averaged sum, is a point on a complex plane that must lie within a convex polygon whose vertices are the eigenvalues of $r_A$, see Fig.~\ref{fig:geometrical_solution}.
We can now distinguish two possibilities, depending on whether the polygon covers the origin.
If it does not, as in the left panel of Fig.~\ref{fig:geometrical_solution}, the energy spectrum has a lower bound $\epsilon_\textrm{min}$ determined by the minimum distance of the polygon from the origin:
\begin{equation}
\label{eq:lower_bound2}
\epsilon_\textrm{min}=\Delta\,\min_{ij} \,[\cos\tfrac{1}{2}(\phi_i-\phi_j)\,]\,.
\end{equation}

\paragraph{Occurrence and physical meaning of zero energy solutions.}

On the other hand, if the polygon covers the origin, as in the right panel of Fig.~\ref{fig:geometrical_solution}, then a zero energy solution $\epsilon=0$ is allowed.
If we order $\phi_1\leq \phi_2 \leq \dots \leq \phi_m$ and introduce phase differences between closest phases $\theta_i=\phi_{i+1}-\phi_i\,\in (-\pi, \pi]$, this happens if
\begin{equation}\label{eq:winding}
\sum_{i=1}^{m} \theta_i=2\pi\,.
\end{equation}
We call the situation of a non-zero winding of the superconducting phases in the leads a ``discrete vortex''.

\paragraph{Crossings are in fact topological transitions.}

Zero energy solutions are doubly degenerate and identify Andreev level crossings at Fermi energy.
These crossings can be seen as topological transitions protected by fermion parity conservation.
At the two sides of the gap closing point, the Pfaffian of the Hamiltonian has opposite signs, which means that energy of a single Andreev state must vanish at the transition point.
Due to the number of modes in the leads being even, crossings can only occur in pairs when advancing any phase by $2\pi$ and for this reason the resulting ground state energy is $2\pi$-periodic.
Conversely, the $4\pi$-periodic Josephson effect, a hallmark of topological superconductivity \cite{Kitaev2001,Kwon2003,Fu2009},  requires an odd number of crossings in a $2\pi$ phase interval, the fermion parity anomaly.

\paragraph{In case you have not noticed yet, the results are completely general.}

We note that the results \eqref{eq:lower_bound2} and \eqref{eq:winding} are quite general: they hold for any number of leads and for arbitrary scattering matrices of the junction.
Hence they are independent of any microscopic detail.
The lower bound of Eq.~\eqref{eq:lower_bound2} is only valid in the short junction limit, while Eq.~\eqref{eq:winding} applies in fact to absolutely any Josephson junction since it is a Fermi level property.

\subsection{multiterminal Josephson junction in the quantum spin Hall regime}
\label{sec:qsh-optimal}

\paragraph{A QSH bar in the topological regime provides the `optimal case'.}

We observe that the lower bound \eqref{eq:lower_bound2} corresponds to the spectrum of a fully transmitted mode connecting two leads.
This scenario can be realized in a quantum spin Hall insulator \cite{Kane2005,Kane2005a,Bernevig2006,Konig2007}.
In this case the Andreev spectrum will depend only on the phase differences between adjacent leads that are connected by topologically protected helical edge states.
In fact, a straightforward generalization of the two-terminal junction of Ref.~\onlinecite{Fu2009} yields the Andreev spectrum
\begin{equation}\label{eq:QSH_andreev_levels}
\epsilon_i=\pm\Delta \cos\,[\tfrac{1}{2}(\phi_{i+1}-\phi_i)]\,,\;\;i=1\,\dots,m\,.
\end{equation}
In a QSH insulator a crossing at zero energy occurs whenever one of the phase differences $\phi_{i+1}-\phi_i=\pi$ [see also the bottom left panel of Fig.~\eqref{fig:crossings}]. For a junction with three leads, this maximizes the region of the phase space with odd ground state fermion parity.

\section{Applications}
\label{sec:numerics}

\paragraph{We consider three physical systems}

We now verify the results of the previous Section applied to junctions with three superconducting leads made in different physical systems.
The physical systems that we study are: (i) chaotic quantum dots with random scattering matrices $s$ uniformly sampled \cite{Mezzadri2007} from the circular symplectic ensemble, (ii) quantum dots made out of a quantum well with Rashba spin-orbit coupling, (iii) quantum dots made out of a quantum spin Hall insulator. In the latter two systems we obtain the scattering matrix numerically using a tight-binding simulation.
We refer to these three systems as `RMT', `Rashba' or `QSH' for brevity.

\paragraph{Rashba Hamiltonian}

The Rashba Hamiltonian describing a 2D electron gas is given by
\begin{equation}\label{eq:rashba_ham}
H=\frac{\vt{p}^2}{2m}+\alpha\,(p_x\sigma_y-p_y\sigma_x)-\mu + V(\vt{r})\,,
\end{equation}
with $\vt{p}=(p_x, p_y)$ the momentum operator, $\sigma_x$ and $\sigma_y$ the spin Pauli matrices, $\alpha$ the strength of the spin-orbit coupling, and $\mu$ the chemical potential.
The disordered electrostatic potential is given by $V(\vt{r})$. This Hamiltonian has time-reversal symmetry with operator $\Theta = i\sigma_y$.

\paragraph{BHZ Hamiltonian}

The quantum spin Hall insulator is described by the Bernevig-Hughes-Zhang model \cite{Bernevig2006}, applicable to HgTe/HgCdTe and InAs/GaAs/AlSb quantum wells. For the numerical simulations, we use the extended model of Ref.~\onlinecite{Liu2008}  (see Appendix \ref{app:BHZ}), which includes spin-orbit coupling contributions due to bulk inversion asymmetry and structural inversion asymmetry, and the material parameters reported in Ref.~\onlinecite{Liu2013}.

\paragraph{Some details on numerics}

To extract the three-terminal scattering matrices of the normal state, we discretize the two models on a square lattice with lattice constant $a$.
We adopt the circular dot geometry shown in Fig.~\ref{fig:dot}, with a radius $R=20 a$ and three leads of width $R$.
We consider the electrostatic disorder $V(\vt{r})$ to be uncorrelated and uniformly distributed in an interval $[-u, u]$.
After obtaining the scattering matrix of the junction we use a gauge with $\phi_3=0$ and solve the eigenvalue problem \eqref{eq:eig_equation} as a function of the remaining two phases $\phi_1, \phi_2$.
We perform the numerical simulations using the Kwant code \cite{Groth2014}.
The scripts with the source code are available online as ancillary files for this preprint.

\subsection{Splitting of Kramers degeneracy}
\label{sec:kramers-splitting}

\begin{figure}[htb]
\centerline{\includegraphics[width=\linewidth]{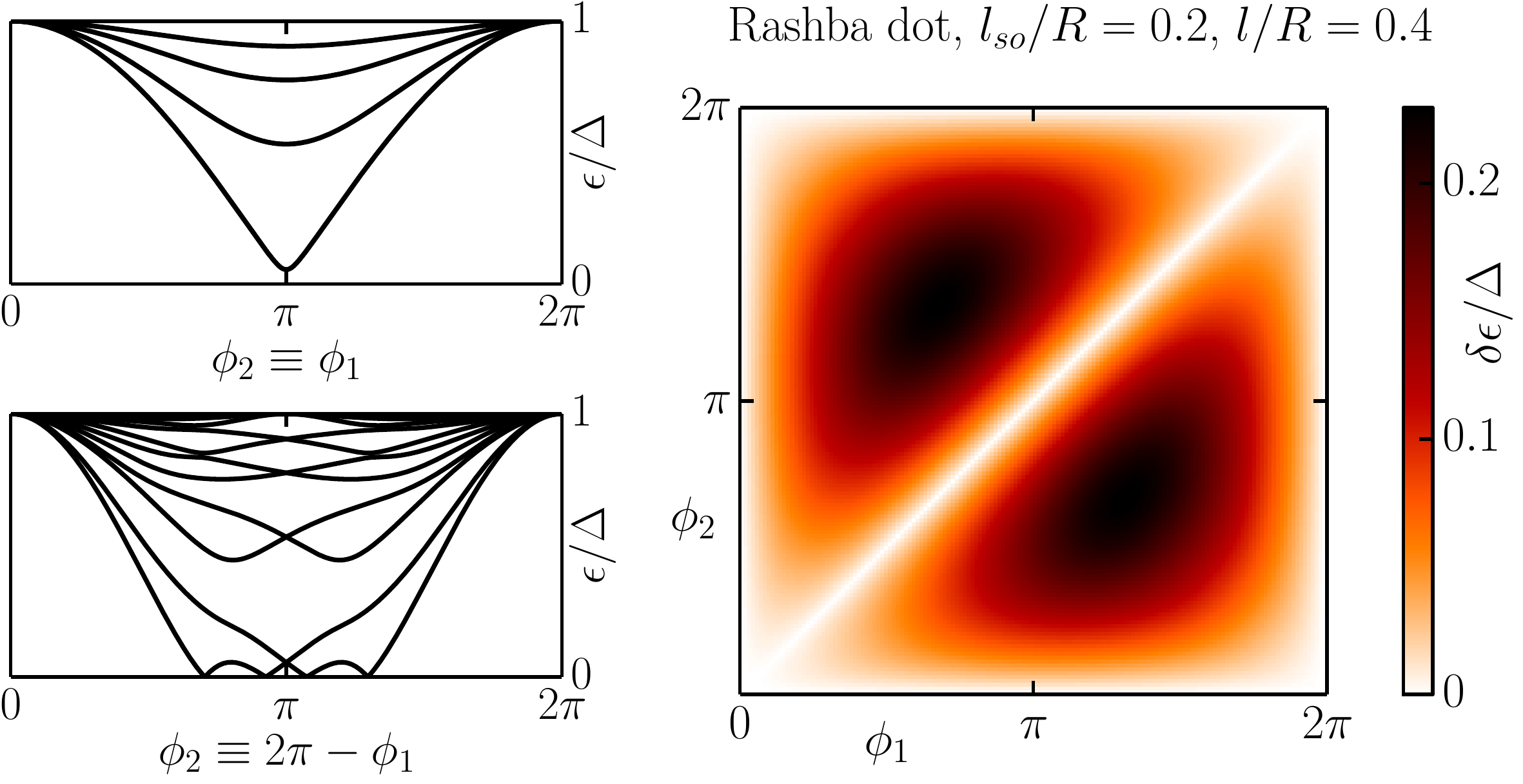}}
\caption{\emph{Left}: Phase dependence of the Andreev levels of a Rashba dot with $\mu=1/4ma$ for $\phi_2=\phi_1$ (top) and $\phi_2=2\pi-\phi_1$ (bottom).
Kramers degeneracy is present in the top panel (since one of the phase differences is zero), but not in the bottom panel.
\emph{Right}: energy difference $\delta \epsilon$ between the two lowest Andreev levels in a Rashba dot  averaged over $10^2$ values of $\mu \in [0,1/2ma]$ for a fixed disorder configuration.}
\label{fig:Kramers_splitting}
\end{figure}

\paragraph{The first thing to look at is the (absence of) Kramers degeneracy.}

The first property we study is the splitting of the Andreev levels.
The two-fold degenerate two-terminal junction spectrum of Eq.~\eqref{eq:2terminal_spectrum} should be recovered whenever any two out of three phase differences are equal, i.e. when either $\phi_1=\phi_2$,  $\phi_1=0$, or $\phi_2=0$.
Away from this limit, we expect deviation from the two-terminal case and a finite splitting of the Kramers doublets.

\paragraph{Brief description of the typical energy spectra.}

A comparison of two typical energy spectra computed for a Rashba dot is shown in the left panel of Fig.~\ref{fig:Kramers_splitting} and confirms our expectations.
To consider the experimentally relevant situation we choose spin-orbit interaction strength $\alpha$ and the disorder strength $u$ such that the spin-orbit length $l_{\rm so} \equiv (m\alpha)^{-1}$ and the mean free path $l \equiv 6\,(mau^2)^{-1}\,\sqrt{\mu/2ma}$ are both smaller than $R$, but have the same order of magnitude.
We first confirm that when $\phi_1=\phi_2$ the spectrum consists of Kramers doublets with the energies given by Eq. \eqref{eq:2terminal_spectrum}.
On the other hand, when the two phases are opposite, $\phi_2=2\pi-\phi_1$, the Kramers pairs of Andreev levels have different energies, except for the time-reversal invariant points $(\phi_1,\phi_2)= 0\!\mod 2\pi$.
One can also notice the presence of  Andreev levels crossings at zero-energy.

\paragraph{We quantify the splitting by looking at the lowest Kramers doublet.}

To quantify the observed splitting of Kramers degeneracy, we consider the energy difference $\delta \epsilon$ between the two levels belonging to the lowest Kramers doublet.
These two levels are of particular interest since they correspond to the most transparent transport channels and their energies are most sensitive to the phase differences.
In the right panel of Fig.~\ref{fig:Kramers_splitting} the splitting $\delta \epsilon$ is computed for a Rashba dot, averaged over different values of $\mu$ in the dot.
It is zero in the two-terminal limit and rises up to $\delta\epsilon\sim 0.2\, \Delta$ away from it.
Hence, Fig.~\ref{fig:Kramers_splitting} confirms our conclusions that Kramers pairs of Andreev levels can be split by an energy of an order $\Delta$ solely by varying the superconducting phases.
The maximal possible splitting is limited by level repulsion, and as expected, we also find that $\delta\epsilon$ is inversely proportional to the total number of Kramers doublets present in the spectrum.

\subsection{Andreev level crossings at zero energy}

\begin{figure}[htb]
\centerline{\includegraphics[width=\linewidth]{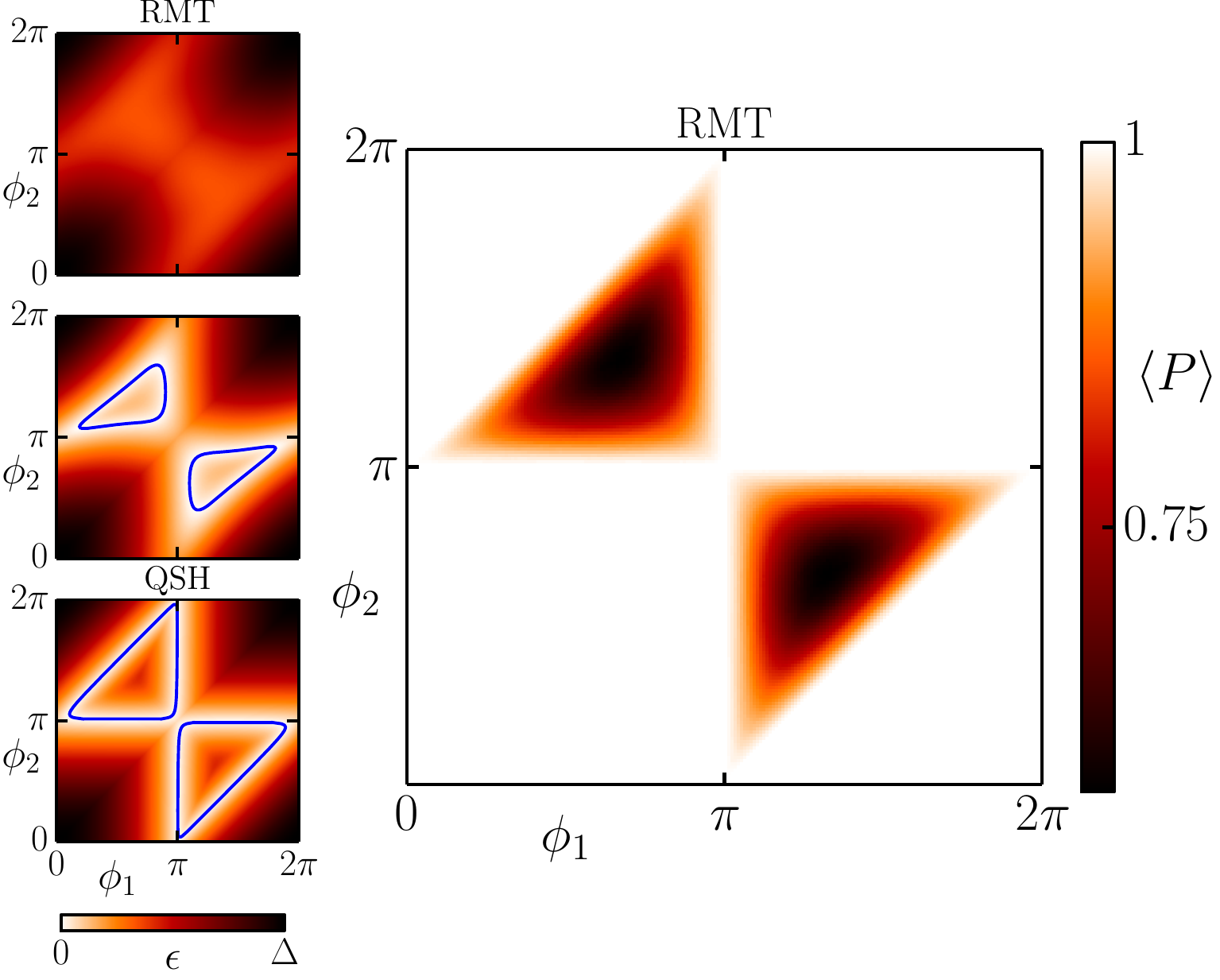}}
\caption{\emph{Left:} Examples of the minimum energy $\epsilon$ of an Andreev bound state as a function of $(\phi_1, \phi_2)$.
The first two examples are calculated using random scattering matrices, with and without with zero-energy crossings.
The positions of the crossings are found numerically using method of App.~\ref{app:gen_eig_eq}, and are marked in blue.
They form closed curves encircling domains of odd ground state fermion parity.
The third example is for a QSH dot in the non-trivial phase, so that the fermion parity switch appears almost exactly at the boundary of the allowed zone.
\emph{Right:} ground state fermion parity $\ev{P}$ averaged over $10^4$ random matrices of size $n=6$, showing that fermion parity may only be odd only if the discrete vortex condition \eqref{eq:winding} is fulfilled.}\label{fig:crossings}
\end{figure}

\paragraph{We proved that crossings can occur, and they do fairly frequently, unlike in two-terminals.}

By checking the Andreev level spectra of different quantum dots, we find that zero-energy crossings indeed occur for some scattering regions, as shown in the left panel of Fig.~\ref{fig:crossings}.
A simulation of a QSH dot\footnote{We use the parameters for an InAs/GaAs/AlSb quantum well, with layer thickness of $10$ nm for both GaSb and InAs \cite{Liu2013}, in a dot with radius $R\simeq 200$ nm, onsite disorder strength $u=25$ meV, and $\mu=0$.} also confirms the conclusion of Sec.~\ref{sec:qsh-optimal} that quantum spin Hall insulators maximize the area in the phase space where the ground state fermion parity is odd.
This behavior is in contrast with that of two-terminal setups, where Eq.~\eqref{eq:2terminal_spectrum} dictates that a Andreev level crossing at zero-energy may only occur in a time-reversal invariant system in the presence of a perfectly transmitted mode.
The stringent requirement of perfect transparency is removed in a multiterminal setup.

\paragraph{We check our predictions for crossings.}

In Section \ref{sec:lower_bound} we proved that zero-energy crossings occur only if a discrete vortex is present at the junction.
For a more systematic study of the occurrence of the zero-energy crossings, we compute the average ground state fermion parity $\ev{P}$ as a function of $\phi_1$ and $\phi_2$ using RMT, with the results shown in the right panel of Fig.~\eqref{fig:crossings}.
The figure shows that the parity deviates from the even value, $\ev{P}=1$, in exact agreement with the vortex condition, Eq.~\eqref{eq:winding}.

\subsection{Density of states}

\paragraph{The properties of the whole spectrum (hard gap vs spectral peak) are strongly influenced by the zero-energy crossing condition.}

\begin{figure}[htb]
\centerline{\includegraphics[width=0.9\linewidth]{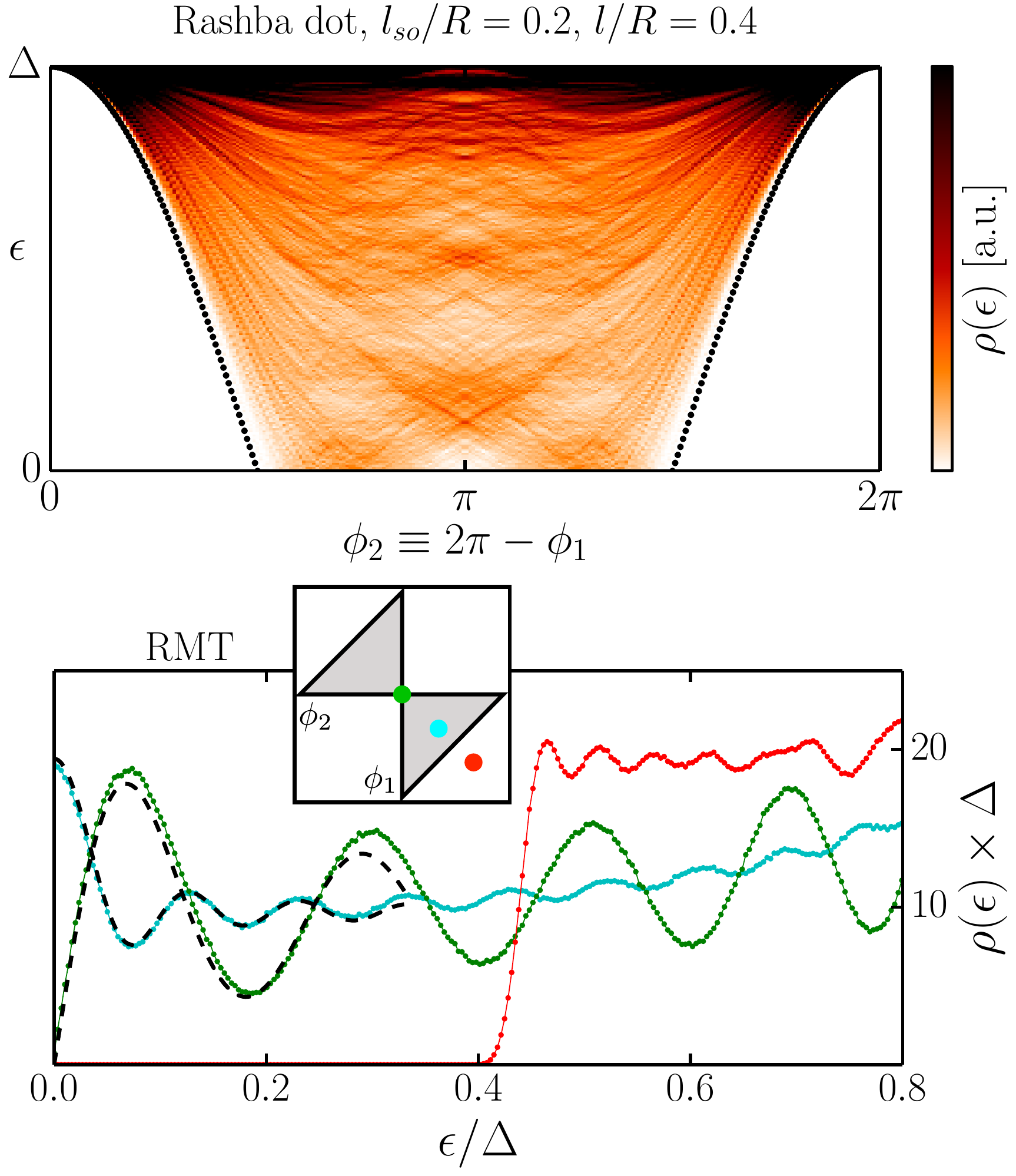}}
\caption{\emph{Top:}  density of states $\rho(\epsilon)$ of a Rashba dot, computed along the diagonal $\phi_2=2\pi-\phi_1$ and averaged over 10$^3$ values of $\mu \in [0,1/2ma]$ for a single disorder realization.
Spin-orbit coupling $\alpha$ and disorder strength $u$ are the same as in Fig.~\ref{fig:Kramers_splitting}.
The dotted line shows the lower bound on the Andreev state energy \eqref{eq:lower_bound2}.
\emph{Bottom:} Density of states obtained from $10^6$ random scattering matrices with 10 modes per lead,  computed for the three different values of $(\phi_1, \phi_2)$ shown in the inset: in the gapped region [red, $(3\pi/4, \pi/4)$], in presence of a discrete vortex [blue, $(4\pi/3, 2\pi/3)$], and at the time-reversal invariant point [green, $(\pi, \pi)$].
The black dashed lines are fits of Eqs. \eqref{eq:DOS_D} and \eqref{eq:DOS_DIII}, with a single free parameter $\delta$.}
\label{fig:dos}
\end{figure}

We now study the properties of the complete Andreev spectrum. In the top panel of Fig.~\ref{fig:dos} we show the subgap density of states $\rho(\epsilon)$ of a Rashba dot, obtained for a single disorder realization while averaging over different values of $\mu$ in the dot.
We observe several features of this density of states.
First, when zero-energy crossings are forbidden an energy gap is present in the spectrum, in agreement with the lower bound of Eq.~\eqref{eq:lower_bound2}.
Second, when crossings are allowed, a spectral peak develops at zero energy.
Finally, at the time-reversal symmetric point $(\phi_1,\phi_2)=(\pi,\pi)$ there is no hard gap in the spectrum but the density of states vanishes at zero energy.

\paragraph{RMT predictions for the density of states}

The latter two features are explained by the random matrix theory of chaotic Andreev dots.
The presence of a spectral peak at zero energy is expected in a chaotic superconducting dot with broken spin rotation and time-reversal symmetries (symmetry class D).
In this case, the expected density of states profile is given by \cite{Mehta,Altland1997,Ivanov2002,Bagrets2012}:
\begin{equation}\label{eq:DOS_D}
\rho(\epsilon)=\delta^{-1}\left[1+\sin(x)/x\right]\,,
\end{equation}
with $x=2\pi \epsilon/\delta$, and  $\delta$ the average level spacing at the Fermi level.
At the time-reversal symmetric point $(\pi, \pi)$ the junction has the symmetry class DIII.
In this case we expect the density of states to vanish at the Fermi level \cite{Nagao1993,Altland1997,Ivanov2002}, with profile
\begin{equation}\label{eq:DOS_DIII}
\rho(\epsilon)=\delta^{-1}\left[\pi^2 x (J'_1(x)J_0(x)+J_1^2(x))+\pi J_1(x)\right]\,,
\end{equation}
where $J_0$ and $J_1$ are Bessel functions of the first kind.

\paragraph{Comparison with RMT predictions}

These corrections to the density of states near zero energy can be observed in our system more clearly by computing the density of states from RMT, see the bottom panel of Fig.~\ref{fig:dos}.
There we compare the density of states at the center of the ``discrete vortex'' $(\phi_1, \phi_2)=(2\pi/3, \pi/3)$ and at the time-reversal symmetric point $(\phi_1, \phi_2)=(\pi, \pi)$ to Eqs. \eqref{eq:DOS_D} and \eqref{eq:DOS_DIII} respectively, using $\delta$ as a fitting parameter.
We find that close to the Fermi level the density profiles are in a good agreement with random matrix theory predictions.
This result is the final confirmation that in a multiterminal short Josephson junction all the consequences of the time-reversal symmetry present in the normal state are removed in the superconducting state by the phase differences.

\subsection{Effect of finite junction size}\label{sec:finite_size}

\begin{figure}[tb]
\centerline{\includegraphics[width=0.9\linewidth]{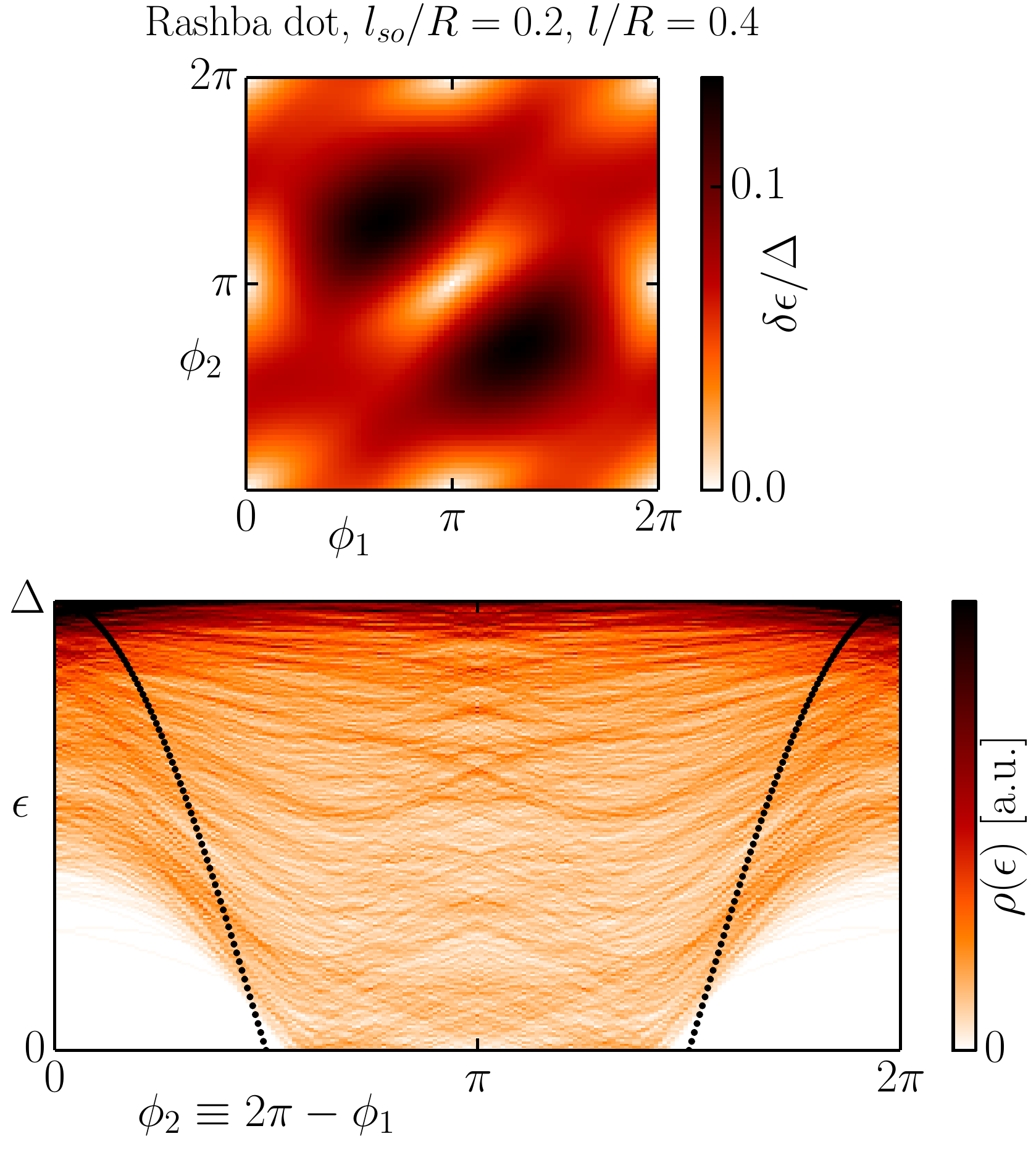}}
\caption{Spectral properties of a three-terminal junction made in a Rashba dot and with finite $\Delta\!=\!0.01 / 2ma$, showing the effect of an increased size of the junction. All other parameters are as in Fig.~\ref{fig:Kramers_splitting}. \emph{Top:} energy difference $\delta \epsilon$ between the two lowest Andreev levels, averaged over 10 values of $\mu \in [0, 1/2ma]$. \emph{Bottom:} density of states of the junction, obtained by averaging over 200 values of $\mu \in [0, 1/2ma]$, for a single disorder configuration and a fixed value of $\mu$ in the three arms of the junction. Black dots are the lower bound \eqref{eq:lower_bound2}, which is valid in the limit $\Delta/E_T\to0$.} \label{fig:finite_size_junction}
\end{figure}

\paragraph{We looked at longer junctions by diagonalizing BdG Hamiltonian.}

Most of our results are applicable in the short junction limit. If the size of the junction is increased, the short junction approximation of Eq.~\eqref{eq:short_junction_condition} gradually loses its validity.
We now consider the corrections to the short junction limit.
In order to do so we include the superconducting pairing explicitly in the Hamiltonian, rather than as a boundary condition for the scattering problem.
We therefore compute the subgap energy spectrum by diagonalizing the Bogoliubov-de Gennes Hamiltonian
\begin{equation}\label{eq:BdG_ham}
H_\textrm{BdG} =
\begin{pmatrix}
  H & \Delta(\vt{r}) \\
  \Delta^*(\vt{r}) & -H
\end{pmatrix},
\end{equation}
where $H$ is the Rashba Hamiltonian \eqref{eq:rashba_ham}.
We apply $H$ to the geometry of Fig.~\ref{fig:dot}, with $\Delta(\vt{r}) = 0$ in the central region and $\Delta(\vt{r}) = \Delta \exp(i\phi_i)$ in the three leads.
We consider finite length leads, interrupted at a distance $L \gtrsim \xi$ away from the junction.

\paragraph{We compare the results with those of a short junction}

In Fig.~\ref{fig:finite_size_junction} we show the results for a junction with $\Delta\!=\!0.01/2ma$, and all other parameters the same as in Sec.~\ref{sec:kramers-splitting}.
As expected, the subgap level spacing and hence the energy splitting of Kramers pairs are reduced in a longer junction.
In particular, the energy splitting of Kramers pair remains finite when two phases in the leads are equal and it only vanishes at time-reversal invariant points.
The lower bound \eqref{eq:lower_bound2} on the energy gap ceases to be valid, as can be seen already from the presence of subgap states at zero phase difference.
Nevertheless, in agreement with our expectations, the vortex condition \eqref{eq:winding} for a zero-energy crossing remains valid.

\section{Conclusions and discussion}\label{sec:discussion}

\paragraph{We offer a new way of electron manipulation}

In conclusion, we have introduced a new method of manipulation of single electron states, which relies solely on applying the superconducting phase differences.
This approach has several advantages over the standard ways that rely on the direct application of magnetic fields.
It allows one to manipulate electron spin locally both in space and time, and to implement long range spin-spin coupling by using inductive coupling of the supercurrents.
Finally, it is not disruptive to superconductivity, making it ideal to apply to hybrid devices.

\paragraph{Specifically we find Kramers and gap closing}

 We demonstrated that, unlike in two terminal Josephson junctions, superconducting phase difference can induce splitting of the Kramers degeneracy in the spectrum comparable to the superconducting gap when more than two superconducting leads are used.
We proved that there is a universal lower bound on the induced gap in the junction, which only depends on the phases of different terminals.
This lower bound vanishes when the phases of the superconducting leads form a discrete vortex.
In that case the ground state fermion parity is allowed to become odd, so that the junction traps an extra fermion in its ground state.

\paragraph{Our findings can be tested directly by tunneling or microwave spectroscopy in a variety of setups}

Our findings can be directly tested experimentally using tunneling spectroscopy.
This requires adding an extra normal or superconducting lead weakly coupled to the scattering region, and performing voltage bias conductance measurements.
The Andreev excitation spectrum of a Josephson junction has also been studied experimentally using microwave absorption spectroscopy \cite{Bretheau2013,Bretheau2014} or measuring switching current probabilities \cite{Zgirski2011,Bretheau2013a}.
Either of these two methods will likewise permit to test our predictions, since both methods are equally applicable to multiterminal junctions.

\paragraph{multiterminal JJs can be made in many materials}

We expect our results to be testable for junctions defined in any material with a sufficiently strong spin-orbit interaction.
Our method of breaking Kramers degeneracy works best in materials with low effective electron mass, since that ensures large normal level spacing.
For instance, for an InAs quantum dot with a radius $R\simeq 100$ nm we estimate a level spacing $\delta_0\simeq \hbar^2\pi^2/8m_{\rm eff}R^2= 0.5$ meV in the normal state, thus making the short junction limit $\Delta\ll n\delta_0$ within easy reach in the case of aluminum contacts.
In addition to the natural candidates such as InAs, InSb quantum wells, or quantum spin Hall insulators, the recently discovered InSb nanocrosses \cite{Plissard2013} make a promising candidate for observing the physics of multiterminal SNS junction.
Conventional metallic SNS junctions would not show the effects of time-reversal symmetry breaking due to the extremely small level spacing.
However, superconducting break junctions \cite{Zgirski2011} could potentially permit the implementation of multiterminal geometries involving a very small number of modes with a large level spacing.

\paragraph{In strong coupling there are also transport signatures.}

There is an entirely different aspect of broken time-reversal and spin rotation symmetries in mesoscopic systems, which is beyond the scope of our investigation, but which can also be studied using our methods.
If the scattering region is additionally strongly coupled to a normal lead, a persistent zero-bias peak in the Andreev conductance is formed \cite{Pikulin2012,Pikulin2012a,Mi2014}.
In our case, we expect such a peak to develop in the presence of a discrete vortex, and to disappear in its absence.

\paragraph{They are very close to recent experimental developments: detection of Andreev bound states\dots}

Another venue of further investigation is to study the quantum nature of the Andreev bound states.
Trapping a single Bogoliubov quasiparticle in a Josephson junction is a promising way to isolate and manipulate a spin degree of freedom - a superconducting spin qubit \cite{Chtchelkatchev2003,Padurariu2010,Padurariu2012}.
A spin-$\tfrac{1}{2}$ state in a Josephson junction is expected to be very stable at low temperatures, due to the energy gap of the superconductor.
These long-lived odd states have been recently observed via switching current measurements in superconducting point contacts \cite{Zgirski2011,Bretheau2013a,Olivares2013}.
The advantage of using multiterminal Josephson junctions for such qubits is that the presence of several tunable phase differences makes it possible to implement universal quantum manipulation exclusively by inductive means.

\paragraph{\dots and topological superconductivity and Majoranas.}

Finally, our discovery provides a better way to creating Majorana bound states in superconductor-semiconductor hybrid systems, a focus of an active experimental search  \cite{Mourik2012,Das2012,Knez2012,Chang2013,Lee2012,Lee2014,Hart2013}.
The complication that arises in many experiments is that magnetic field required to induce a non-trivial gap in the semiconductor is too strong and spoils the properties of the superconductor.
Using superconducting phases as a means of breaking time reversal symmetry and Kramers degeneracy would allow one to reach the same goal without any detrimental effect on the superconductor.
Potentially it would even allow one to use aluminum, which forms high quality contacts with semiconductors and is the simplest superconducting material to use in fabrication, and whose application to Majoranas was so far limited by its extremely small critical field.
One promising use of our method for creation of Majoranas is to combine multiple superconducting leads with an engineered Kitaev chain geometry of Refs.~\onlinecite{Romito2012,Sau2012,Fulga2013}.

\acknowledgments

We thank T. Gatanov for his contribution to finding the proof of the condition~\eqref{eq:lower_bound2} and C. W. J. Beenakker for discussions and useful comments on the manuscript. This research was supported by the Foundation for Fundamental Research on Matter (FOM), the Netherlands Organization for Scientific Research (NWO/OCW), an ERC Synergy Grant, and the China Scholarship Council.

\bibliography{trijj}

\appendix

\section{Occurrence of a zero-energy crossing as a generalized eigenvalue problem}\label{app:gen_eig_eq}

Given the scattering matrices $s$ and $r_A$, it is possible to determine whether zero-energy solutions exist in the $(\phi_1, \phi_2)$ plane without solving for the spectrum. To do so, we can recast Eq.~\eqref{eq:eig_equation} at $\epsilon=0$ as a generalized eigenvalue problem of the form
\begin{equation}\label{eq:gen_eig_prob}
X\,\Psi^e_\textrm{in}=\e^{-i\phi_1}\,Y\,\Psi^e_\textrm{in}
\end{equation}
We give the explicit form of $X$ and $Y$ in the case of three leads. If $s$ has the following block structure,
\begin{equation}
s=
\begin{pmatrix}
  r_{11} & t_{12} & t_{13} \\
  -t_{12}^T & r_{22} & t_{23} \\
  -t_{13}^T & -t_{23}^T & r_{33}
\end{pmatrix}\,,
\end{equation}
then $X$ and $Y$ are given by
\begin{align}\notag
&X=
\begin{pmatrix}
  0 & -\e^{-i\phi_2} t_{12} & -t_{13} \\
  \e^{-i\phi_2} t_{12}^T & 2\e^{-i\phi_2} r_{22} & -\left(1+\e^{-i\phi_2}\right)\,t_{23} \\
  t_{13}^T & \left(1+\e^{-i\phi_2}\right)\,t_{23}^T & 2r_{33}
\end{pmatrix}\,,\\
&Y=
\begin{pmatrix}
  2\,r_{11} & t_{12} & t_{13} \\
  -t_{12}^T & 0 & 0 \\
  -t_{13}^T & 0 & 0
\end{pmatrix}\,.
\end{align}
The existence of a zero-energy crossing at the position $(\phi_1, \phi_2)$ can then be determined numerically by checking that Eq.~\eqref{eq:gen_eig_prob} has eigenvalues with unit norm.

\section{BHZ Hamiltonian}\label{app:BHZ}

The BHZ Hamiltonian describing a 2D quantum spin Hall insulator reads \cite{Liu2008}:
\begin{equation}
H_\textrm{BHZ} = H_{0} + H_\textrm{BIA} +H_\textrm{SIA}+V(\vt{r})\,,
\end{equation}
with $V(\vt{r})$ the electrostatic disorder, and
\begin{subequations}
\begin{align}
H_0  &=
\begin{pmatrix}
  h(\vt{p}) & 0 \\
  0 & h^*(-\vt{p})
\end{pmatrix}\,, \\
H_\textrm{BIA}&=
\begin{pmatrix}
  0 & 0 & \Delta_{e}p_{+} & - \Delta_{z}\\
  0 & 0 & \Delta_z & \Delta_{h}p_{-} \\
  \Delta_{e} p_{-} & \Delta_z & 0 & 0\\
  -\Delta_z & \Delta_{h}p_{+} & 0 & 0
\end{pmatrix}\,, \\
H_\textrm{SIA} &=
\begin{pmatrix}
  0 & 0 & i\xi_{e}p_{-} & 0 \\
  0 & 0 & 0 & 0 \\
  -i\xi^{*}_{e}p_{+} & 0 & 0 & 0 \\
  0 & 0 & 0 & 0
\end{pmatrix}\,,
\end{align}
\end{subequations}
and
\begin{equation*}
h(\vt{p})=(C - D \vt{p}^2) \sigma_0 + A(p_x\sigma_x - p_y\sigma_y) +(M - B\vt{p}^2)\sigma_z\,.
\end{equation*}
Here, $\sigma$ are the Pauli matrices in orbital space, $\vt{p}$ is the momentum operator, and $p_\pm=p_x\pm ip_y$. The system is in a topologically nontrivial phase whenever $M < 0$.

\end{document}